\documentclass[prl,twocolumn,showpacs,groupedaddress]{revtex4}
\usepackage[dvips]{epsfig}

\begin{document}
\title{Nature of the spin dynamics and 1/3 magnetization plateau in azurite}

\author{K.C. Rule$^{1}$, A.U.B. Wolter$^{1}$, S. S\"ullow$^{2}$, D.A. Tennant$^{1}$, A. Br\"uhl$^{3}$, S. K\"{o}hler$^{3}$, B. Wolf$^{3}$, M. Lang$^{3}$, J. Schreuer$^{4}$}

\affiliation{$^{1}$Hahn-Meitner-Institut GmbH, D-14109 Berlin, Germany}
\affiliation{$^{2}$Inst. f\"ur Physik der Kondensierten Materie, TU Braunschweig, D-38106 Braunschweig, Germany}
\affiliation{$^{3}$Physikalisches Institut, J.W. Goethe-Universit\"at Frankfurt, D-60438 Frankfurt(M), Germany}
\affiliation{$^{4}$Ruhr-Universit\"at Bochum, Bochum, Germany}

\date{\today}

\begin{abstract}
We present a specific heat and inelastic neutron scattering study in magnetic fields up into the 1/3 magnetization plateau phase of the diamond chain compound azurite Cu$_3$(CO$_3$)$_2$(OH)$_2$. We establish that the magnetization plateau is a dimer-monomer state, {\it i.e.}, consisting of a chain of $S = 1/2$ monomers, which are separated by $S = 0$ dimers on the diamond chain backbone. The effective spin couplings $J_{mono}/k_B = 10.1(2)$ K and $J_{dimer}/k_B = 1.8(1)$ K are derived from the monomer and dimer dispersions. They are associated to microscopic couplings $J_1/k_B = 1(2)$ K, $J_2/k_B = 55(5)$ K and a ferromagnetic $J_3/k_B = -20(5)$ K, possibly as result of $d_{z^2}$ orbitals in the Cu-O bonds providing the superexchange pathways.
\end{abstract}


\maketitle

Great interest has surrounded the observation of a 1/3 magnetization plateau in azurite Cu$_3$(CO$_3$)$_2$(OH)$_2$ \cite{kikuchi,gu}. This material, famous as a painting pigment of deep-blue colour, has been proposed as a realisation of the exotic diamond-chain Hamiltonian of coupled spin-1/2 moments, written as
\begin{eqnarray}
\hat{H} = J_1 \sum^{N/3}_{j=1} ({\bf S}_{3j-1} {\bf S}_{3j} + {\bf S}_{3j} {\bf S}_{3j+1}) + J_2 \sum^{N/3}_{j=1} {\bf S}_{3j+1} {\bf S}_{3j+2} \nonumber \\ 
+ J_3 \sum^{N/3}_{j=1} ({\bf S}_{3j-2} {\bf S}_{3j} + {\bf S}_{3j} {\bf S}_{3j+2}) - g \mu_B B \sum^{N}_{j=1} S_j^z.
\end{eqnarray}
Here, $J_2$ is the magnetic coupling of the diamond backbone, while $J_1$ and $J_3$ represent the coupling of the monomers along the chain \cite{takano,okamoto,honecker} (Fig.~\ref{fig:fig2}). Depending on the relative coupling strengths $J_1, J_2, J_3$, this model affords a host of exotic phases and quantum phase transitions, including possibly $M = 1/3$ fractionalisation \cite{mueller} or exotic dimer phases \cite{okamoto}. However, determining the magnetic exchange couplings in azurite has proved difficult, yielding controversial results. While a susceptibility $\chi$ study claims $J_1/k_B = 19$ K, $J_2/k_B = 24$ K, $J_3/k_B = 8.6$ K, implying strong frustration \cite{kikuchi}, subsequent numerical studies of $\chi$ dispute this claim, proposing a ferromagnetic (FM) $J_3$, and thus a non-frustrated scenario \cite{gu}.

The general issue underlying these starkly contrasting interpretations of the same experimental data is that of the nature of magnetic coupling in low-dimensional (low-D) quantum magnets. In azurite Cu$_3$(CO$_3$)$_2$(OH)$_2$, the Cu$^{2+}$ ions ($S = 1/2$) are in a square-planar coordination on two inequivalent sites \cite{reddy}. The system has a monoclinic crystal structure (space group $\it P2_1/c$, lattice parameters $a$ = 5.01 \AA , $b$ = 5.85 \AA , $c$ = 10.3 \AA , $\beta = 92.4^{\circ}$ \cite{gattow,belokoneva}), where the Cu$^{2+}$ network is built up by diamond-shaped units arranged in chains running along the $b$ direction. The magnetic exchange pathways for the $J_1$, $J_2$ and $J_3$ couplings are along Cu-O-Cu bonds with angles of 113.7$^{\circ}$, 97$^{\circ}$ and 113.4$^{\circ}$, respectively. Hence, the Goodenough-Kanamori-Anderson rules for superexchange \cite{goodenough} would predict the couplings to be weakly antiferromagnetic (AFM) \cite{note1}.

However, this conclusion is only valid if the magnetic orbitals are dominantly of $d_{x^2 - y^2}$ character. If the Cu orbitals tend to be more $d_{z^2}$ like, higher-order effects come into play, which can give rise to FM exchange \cite{filipetti}. The latter case may become relevant for systems with bond angles away from the limiting cases of 90$^{\circ}$ and 180$^{\circ}$, such as CuO or some molecular magnets \cite{yang}. In fact, recent x-ray diffraction experiments \cite{belokoneva} on azurite indicate a dominant $d_{z^2}$ character of Cu$^{2+}$, making azurite another prominent candidate for such superexchange interactions.

In this Letter we present a detailed study of azurite by means of specific heat and inelastic neutron scattering in zero and applied magnetic field. We investigate the elementary magnetic excitations, in particular the {\bf q} dependence of the spin-excitation spectrum in the plateau phase, and estimate the microscopic coupling constants.

\begin{figure}[!ht]
\begin{center}
\includegraphics[width=0.8\linewidth]{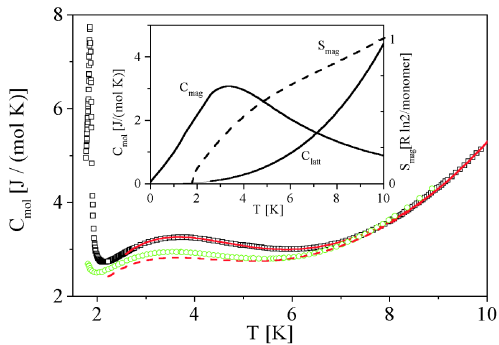}
\end{center}
\caption{(Color online) $C_{p} (T)$ of a single crystal of azurite for $B = 0$ T (squares) and 4 T ($\bigcirc$). The solid (dashed) line represents a fit to the zero-field ($B = 4$ T) data using a magnetic ($C_{mag}$) \cite{kluemper2} and phononic ($C_{lat}$) contribution. The inset shows $C_{mag}(T)$ at $B = 0$ T, $C_{lat}$ and the magnetic entropy $S_{mag}$ derived by integrating $(C_{p} (T) - C_{lat} (T))/T$.} \label{fig:fig2a}
\end{figure}

Specific heat $C_p (T)$ measurements at temperatures 1.6 to 30 K were carried out using an ac-calorimeter \cite{sullivan} on an azurite crystal (mass: 0.36 mg), which was cut from the crystal used for neutron scattering. An external field up to 4 T was oriented in the $ac$-plane, and 65$^{\circ}$ away from the $c$ axis. This orientation corresponds approximately to the easy axis of the AFM phase below $T_N = 1.85$ K \cite{love}. The $C_p (T)$ results are displayed in Fig.~\ref{fig:fig2a} up to 10 K. Beside the AFM transition, indicated by a sharp discontinuity, the zero-field data reveal a broad maximum around 3.7 K. At $B = 4$ T the maximum becomes reduced in size and shifted to lower temperatures.

The starting point of our analysis is the \textit{dimer-monomer} model proposed in Ref.~\cite{okamoto}, with the intradimer coupling constant $J_2$ representing the dominant energy scale. At temperatures $T < 10$ K, considered here for the specific heat, and for $J_2/k_B \gg 10$ K (see below), the magnetic degrees of freedom of azurite are dominated by the chain of spin-1/2 Cu$^{2+}$-monomers, which are antiferromagnetically coupled via the rungs of the diamond backbone \cite{kikuchi}. Using the magnetic specific heat $C_{mag}$ of the AFM $S = 1/2$ Heisenberg chain (AFHC) \cite{kluemper2}, and including a lattice contribution $C_{lat} \propto (T/\Theta_{D})^3$, the zero-field specific heat $C_{p} (T)$ is well fitted down to 2.5 K. In this fit, the magnetic coupling $J_{AFHC}/k_B = 7.0(1)$ K and a Debye temperature $\Theta_{D} = 188$ K were used (solid line in Fig.~\ref{fig:fig2a}). The corresponding magnetic entropy $S_{mag}$, normalized to a Cu$^{2+}$ monomer, reaches the full value $R ln2$ at about 10 K (broken line in the inset of Fig.~\ref{fig:fig2a}). Using the same lattice contribution and $C_{mag} (T,B)$ results calculated for the AFHC in finite field \cite{kluemper2}, this ansatz (broken line in Fig.~\ref{fig:fig2a}) also captures the main features of the experimental data at $B = 4$ T.

Low-temperature ($T = 1.5$ K) inelastic neutron scattering (INS) was used to further probe the character of the magnetic excitations both at zero and high magnetic field. Experiments were carried out at the Berlin Neutron Scattering Center, BENSC, using the V2 cold-neutron triple-axis spectrometer. Constant-{\bf Q}, energy transfer scans in a range 0 - 7.8 meV were performed with incident neutrons fixed at $k_i = 1.55$ and 1.3 \AA$^{-1}$ which gave an energy resolution of 0.15 and 0.11 meV, respectively. A large ($> 15$ g), naturally grown azurite single crystal was mounted with a horizontal $a^* - b^*$ scattering plane, and magnetic fields of up to 14 T applied perpendicular to $b^*$. Susceptibility data, taken on pieces of this crystal, perfectly reproduce the results reported in Ref.~\cite{kikuchi}.

\begin{figure}[!ht]
\begin{center}
\includegraphics[width=1\linewidth]{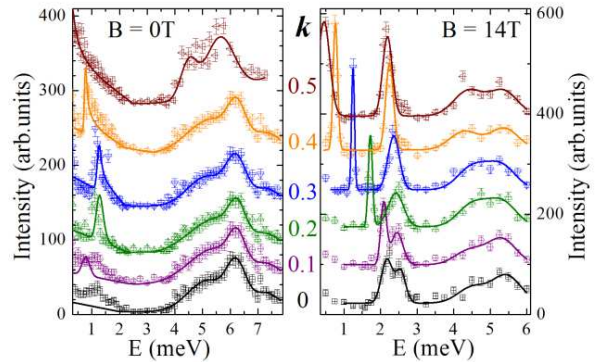}
\end{center}
\caption{(Color online) Energy dependence of the INS spectra of azurite at {\bf q} $= [1 ~ k ~ 0]$, $k = 0 - 0.5$, in zero field (left) and a field of 14 T $\perp b^*$ (right) at $T = 1.5$ K. Lines indicate fits to the data, which are shifted for clarity; for details see text.} \label{fig:fig1}
\end{figure}

\begin{figure}[!ht]
\begin{center}
\includegraphics[width=0.6\linewidth]{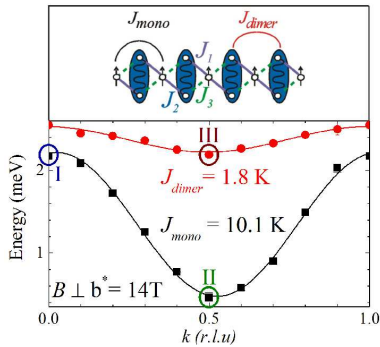}
\end{center}
\caption{(Color online) (top) \textit{Dimer-monomer} model with microscopic couplings $J_{1}$, $J_{2}$, $J_{3}$, and effective couplings $J_{mono}$, $J_{dimer}$, characterizing the plateau phase. (bottom) $E(k)$ dependence along {\bf q} $= [1 ~ k ~ 0]$ of azurite for the two low-lying excitations in $B (\perp b^{*}) = 14$T. Lines are fits to the data, see text. The small asymmetry of the data with respect to $k = 0.5$ reflects a slight misalignment of the crystal. Highlights I, II, III indicate positions investigated for their $B$ dependence in Fig.~\ref{fig:fig3}.}
\label{fig:fig2}
\end{figure}

The essential features of our INS studies are summarized in Fig.~\ref{fig:fig1}. Here, we plot the energy dependence of the scattering intensity for various positions {\bf q} $= [1 ~ k ~ 0]$ along the diamond chain in zero field and in 14 T, {\it i.e.}, in the 1/3 magnetization plateau phase between $B_{c1} = 11$ T and $B_{c2} = 30$ T. Data shown here only cover the range $0 \leq k \leq 0.5$ since measurements up to $k = 1$ reveal symmetric behaviour with respect to $k = 0.5$.

Along {\bf q} $= [1 ~ k ~ 0]$ and in zero magnetic field, there is scattering intensity at energies $<$ 2 meV displaying the $|\sin{\bf q}|$ dependence of an $S = 1/2$ AFHC (left panel Fig. \ref{fig:fig1}). The fits shown in the figure were produced by combining Gaussian line shapes for the features above 2 meV, while those below 2 meV were calculated using a 1-D chain \cite{caux} convolved with instrumental resolution and the copper form factor. The fit yields an effective magnetic coupling strength of $J_{AFHC}/k_B = 9(1)$ K (solid lines $<$ 2 meV). The difference between $J_{AFHC}$ measured from specific heat and inelastic neutron scattering is not well understood and will require additional studies. Perpendicular to the chain, {\it i.e.}, along $[h ~ 0.5 ~ 0]$, within experimental resolution there is no dispersion (not shown), consistent with the 1-D nature of these spin excitations.

Further, our spectra reveal a broad peak structure at energies above 3.5 meV in zero magnetic field (left panel Fig.~\ref{fig:fig1}). As we will argue below, this feature arises from singlet-triplet excitations of the spin dimers on the backbone of the diamond chain structure.

Within the \textit{dimer-monomer} model, the monomer chain becomes fully polarized at the lower critical field $B_{c1}$ of the plateau phase. Correspondingly, a gap opens at the AFM point {\bf q} $= [1 ~ 0.5 ~ 0]$. The excitations are now ferromagnons, yielding scattering intensity with a cosinusodial dispersion (see low-energy peak $0.4 \leq E \leq 2.2$ meV in 14 T; right panel Fig.~\ref{fig:fig1}). By fitting the peak positions using Gaussian profiles as indicated in the plot, we extract the $E(k)$ dependence in Fig.~\ref{fig:fig2} (black squares). The data are parameterized by an expression $E(k) = g_{av} \mu_B B + J_{mono} (cos (2 \pi k ) - 1) + \Delta_{chain}$. With $g_{av} = 2.1$ from ESR measurements \cite{ishii} and $\Delta_{chain} = 0.53$ meV, we obtain the effective magnetic coupling $J_{mono}/k_B = 10.1(2)$ K. Using higher order cosine-terms does not improve the fitting, resulting in 0.2 K as upper boundary for next-nearest-neighbor interactions. 

The broad peak in the spectra above 3.5 meV in Fig. \ref{fig:fig1} is attributed to singlet-triplet excitations within the dimers. In magnetic fields, Zeeman splitting lifts the zero-field degeneracy of the singlet-triplet peak, lowering the $\left| \uparrow \uparrow \right\rangle$ triplet branch in energy and separating it from the broad structure. Accordingly, in $B = 14$ T, a weakly dispersive peak is observed at $\sim 2.2 - 2.5$ meV (right panel Fig.~\ref{fig:fig1}). By fitting the peak positions of the low-lying triplet using Gaussian profiles, we extract the $E(k)$ dependence in Fig.~\ref{fig:fig2} (red circles). It can be parameterized by $E(k) = - g_{av} \mu_B B + J_2 + J_{dimer} cos (2 \pi k ) + \mu_B \tilde{b}$, with $g_{av} = 2.1$ \cite{ishii} and $J_2$ as zero-field singlet-triplet splitting. The parameter $\tilde{b}$ accounts for the internal field shift at the dimer site due to the alignment of nearest-neighbouring monomers. From the width of the dispersion we obtain an effective dimer-dimer coupling $J_{dimer} = 1.8$ K. The other triplet branches, even in highest fields, superimpose and yield a broad distribution of scattering intensity for energies $> 3.5$ meV, thus prohibiting a precise determination of peak positions. As yet, the cause for the peak broadening of these branches is not understood.

\begin{figure}[!ht]
\begin{center}
\includegraphics[width=1\linewidth]{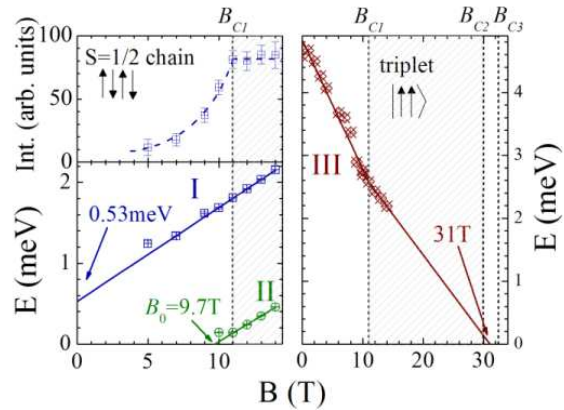}
\end{center}
\caption{(Color online) (left) $B$ dependence of the integrated peak intensity (top) and monomer energy (bottom) at {\bf q} $= [1 ~ 0 ~ 0]$ (I) and $[1 ~ 0.5 ~ 0]$ (II). (right) Energy of the low-lying triplet branch ($\left| \uparrow \uparrow \right\rangle$) at {\bf q} $= [1 ~ 0.5 ~ 0]$ (III) as function of $B$. Hatched region indicates the plateau phase. Solid lines: fits to the data; dashed lines: guides to the eye.} \label{fig:fig3}
\end{figure}

The critical fields of the plateau phase are extracted from a field-dependent study of the monomer and lowest dimer excitations. First, we note that the field dependence of the monomer dispersion should be that of the $S = 1/2$ AFHC. Hence, at {\bf q} $= [1 ~ 0 ~ 0]$ (marker I in Fig.~\ref{fig:fig2}) a (close to) linear-in-field behaviour is expected, which is gapless in zero field \cite{kluemper}. As shown in the lower left panel of Fig.~\ref{fig:fig3}, a linear behaviour is in fact observed experimentally. Further, $B_{c1} = 11$ T is identified in the field dependence of the peak intensity (upper left panel Fig.~\ref{fig:fig3}) as a distinct kink. The intensity is constant above $B_{c1}$, but it falls off rapidly with decreasing field for $B < B_{c1}$. The intensity decrease is due to the first moment sum rule and is related to the magnetization \cite{hohenberg}.

Surprisingly, an extrapolation of the peak position for $B \geq B_{c1}$ to zero field yields a finite gap $\Delta_{chain} = 0.53$ meV. This feature also accounts for the mismatch at low energies between fit and data at {\bf q} $= [1 ~ 0 ~ 0]$ in zero field (left panel Fig.~\ref{fig:fig1}). Unfortunately, a detailed study of this behaviour is hampered by the strong reduction of peak intensity (upper left panel Fig.~\ref{fig:fig3}). Possibly, the AFM ordered state affects the behaviour of the monomer chain, leading to a residual zero-field gap at the nuclear zone centre. Certainly, this issue calls for further investigations.

In addition, we measured the $B$ dependence of the monomer peak at $[1 ~ 0.4 ~ 0]$. Assuming a $cos ( 2 \pi k )$ dependence of $E (k)$, we approximate the field dependence of the $E (k)$ minimum at $[1 ~ 0.5 ~ 0]$ (marker II in Fig. \ref{fig:fig2}), this way accessing the field of gap closure for the ferromagnons \cite{note2}. For $B \geq
B_{c1}$ the expected linear-in-field behaviour of the gap is observed experimentally, with an extrapolated field of gap closure $B_0 = 9.7$ T (lower left panel Fig. \ref{fig:fig3}). We associate the difference between $B_{c1}$ and $B_0$ to the local mean field due to the AFM order. 

Our data suggest that the magnetization plateau reaches from the field $B_{c1}$ of full monomer polarization up to $B_{c2}$, where the low-lying triplet $\left| \uparrow \uparrow \right\rangle$ reaches zero energy. Moreover, Zeeman splitting results in a dependence $\propto B$ for this triplet branch. To test this, we have measured the field dependence of the $\left| \uparrow \uparrow \right\rangle$ branch at the minimum {\bf q} $= [1 ~ 0.5 ~ 0]$ (marker III in Fig. \ref{fig:fig2}; data in Fig. \ref{fig:fig3}, right panel). As expected, for fields $B \geq B_{c1}$ a linear field dependence is observed. An extrapolation to zero energy from the low-lying triplet $\left| \uparrow \uparrow \right\rangle$ at {\bf q} $= [1 ~ 0.5 ~ 0]$ yields a value of 31 T, which closely matches the reported value $B_{c2} = 30$ T as upper critical field of the magnetization plateau. Extrapolating to zero energy for {\bf q} $= [1 ~ 0 ~ 0]$ gives $B_{c3} = 33.6$ T also comparable to the reported value \cite{kikuchi} of 32.5 T for the onset of magnetization saturation to the full moment.

Below $B_{c1}$, for the triplet $\left| \uparrow \uparrow \right\rangle$ at {\bf q} $= [1 ~ 0.5 ~ 0]$ again a behaviour $\propto B$, but with a steeper negative slope, is observed. This reflects the presence of local molecular fields as a result of the AFM aligned monomers. Fitting the data at $B \leq B_{c1}$ yields a zero field singlet-triplet splitting of 4.8(5) meV ($\sim 55(5)$ K). This splitting essentially corresponds to the magnetic coupling $J_2$ on the diamond backbone.

Altogether, our data reveal that the magnetization plateau in azurite arises out of the {\it dimer-monomer} state proposed in Ref.~\cite{okamoto} and indicated in the top panel of Fig.~\ref{fig:fig2}. There is a chain of antiferromagnetically coupled $S = 1/2$ monomers, which are fully polarized in the plateau phase. The coupling between the monomers is provided by the rungs in the diamond units, which carry two spins forming a dimer ($J_2$). Nearest-neighbor dimers are weakly coupled ($J_{dimer}$) to each other.

Within first-order perturbation theory, analytical expressions relate $J_{2}$, $J_{dimer}$ to the microscopic couplings $J_1, J_3$, provided that $J_2 \gg J_1, J_3$ \cite{honecker}. Although in azurite the latter condition might not be strictly fulfilled, our data indicate $J_{2}$ to be the strongest coupling. Employing $J_2$ and $J_{dimer}$ from experiment, we approximate $J_1$ and $J_3$ following Ref.~\cite{honecker}. First, for the dimer coupling it predicts $J_{dimer} \approx (J_1 - J_3)^2 / 4 J_2$. Secondly, the change in slope of the dimer dispersion at $B_{c1}$ (Fig.~\ref{fig:fig3}) arises from a change in local mean fields from AFM to FM. Extrapolating the data at $B > B_{c1}$ to zero field we obtain the difference between FM and AFM polarization $\Delta E \approx ( J_1 + J_3 )/2 + J_{dimer} = -0.8$ meV. Finally, we can use the various critical field expressions such as $B_{c2} \approx (J_2 - J_{dimer} + (J_1 + J_3)/2)/g_{av} \mu_B $ for checking the consistency of our calculations. With this procedure we obtain $J_3/k_B \simeq -20(5)$ K and $J_1/k_B \simeq 1(2)$ K.

Most importantly, our findings indicate that there is a FM coupling in azurite, as was suggested previously \cite{gu}. The $J_3$ path has Cu-O-Cu exchange along one filled and one unfilled orthogonal orbital, with atomic Cu-O distances of 1.947 and 1.989 \AA ~separated by an angle of 113.4$^{\circ}$. By reducing the Anderson superexchange mechanism and with the Cu orbitals of a dominantly $d_{z^2}$ character \cite{belokoneva}, direct exchange and oxygen Hund's rule coupling can yield a substantial ferromagnetic exchange pathway \cite{filipetti}. Further, with $|J_1| \ll |J_2|, |J_3|$ frustration effects will be absent or weak. $J_1$ represents a weak perturbation, and is possibly less important than other factors such as Dzyaloshinskii-Moriya-interaction \cite{kikuchi}.

To conclude, we have studied the distorted diamond chain model system azurite Cu$_3$(CO$_3$)$_2$(OH)$_2$ by means of specific heat and inelastic neutron scattering. The magnetization plateau can be understood within the \textit{dimer-monomer} model from Ref.~\cite{okamoto}. For the microscopic couplings $J_1$, $J_2$ and $J_3$ within the diamond units we find $J_2$ and $J_3$ to be dominant, with $J_3$ being ferromagnetic. Accounting for this type of superexchange ought to be possible along the lines set out in Ref. \cite{filipetti}. Given the limits of our perturbative analytic approach, further calculations, which take the renormalisation of magnetic couplings by quantum fluctuations into account, are needed to fully model the distorted diamond chain azurite Cu$_3$(CO$_3$)$_2$(OH)$_2$ \cite{Mikeska}. 

This work has been supported by the BENSC and the SFB/TRR 49. Fruitful discussions with H. Mikeska, A. Honecker, J. Richter, S. Gro\ss johann and W. Brenig are gratefully acknowledged.


\begin{references}
\bibitem{note1} The $J_2$ bond angle is close to FM exchange, but proven to be AFM by experiment; see Ref. \cite{kikuchi}.
\bibitem{note2} This position was not accessible directly because of the incoherent elastic scattering.
\end{references}
\end{document}